\documentclass[aps,prb,reprint,superscriptaddress,floatfix]{revtex4-2}

\usepackage{amsmath}
\usepackage{amssymb}
\usepackage{graphicx}
\usepackage{bm}
\usepackage{multirow}
\usepackage[colorlinks=true,allcolors=blue]{hyperref}

\begin{document}

\title{$\mu$SR study of time-reversal symmetry constraints and bulk superfluid response in Li$_{0.95}$FeAs}

\author{Rustem Khasanov}
\email{rustem.khasanov@psi.ch}
\affiliation{PSI Center for Neutron and Muon Sciences CNM, 5232 Villigen PSI, Switzerland}
\author{Hubertus Luetkens}
\affiliation{PSI Center for Neutron and Muon Sciences CNM, 5232 Villigen PSI, Switzerland}
\author{Nikolai D. Zhigadlo}
\affiliation{CrystMat Company, CH-8037 Zurich, Switzerland}

\date{April 2026}

\begin{abstract}
We report zero-field (ZF) and transverse-field (TF) muon-spin rotation/relaxation ($\mu$SR) measurements on superconducting Li$_{0.95}$FeAs ($T_{\rm c}\simeq16.0$~K) grown by a high-pressure self-flux method. The ZF-$\mu$SR data show no detectable change of the electronic relaxation rate on cooling through $T_{\rm c}$, providing no evidence for time-reversal-symmetry breaking in the superconducting state. TF-$\mu$SR measurements reveal a well-developed vortex response with strong flux pinning and a negligible nonsuperconducting contribution, confirming that superconductivity is a bulk property of the sample. From the second moment of the internal field distribution we determine a low-temperature in-plane magnetic penetration depth $\lambda_{ab}= 245(15)$~nm. The temperature dependence of the normalized superfluid density is well described by an effective two-gap model with $\Delta_1 = 2.0(2)$~meV and $\Delta_2 = 0.7(2)$~meV. A quantitative comparison with ARPES-based band weights shows that the $\mu$SR response is dominated by the Fermi-surface sheets carrying the intermediate and small superconducting gaps, whereas the band hosting the largest gap contributes only about 3\% to the total superfluid density and is therefore not resolved in the present analysis. Taken together, these results establish Li$_{0.95}$FeAs as a bulk multigap superconductor without detectable time-reversal-symmetry breaking and show how $\mu$SR reconciles the gap scales reported by bulk and surface-sensitive probes in this multiband system.
\end{abstract}

\maketitle

\section{Introduction}

LiFeAs is one of the most intriguing members of the Fe-based superconductors (FeSCs) \cite{wang2008, tapp2008}. Unlike many FeSCs, where superconductivity emerges only after chemical doping suppresses magnetic order, LiFeAs is intrinsically superconducting in stoichiometric form, without the need for external chemical substitution or pressure \cite{tapp2008,pitcher2010,wissmann2022}. Moreover, it does not show long-range magnetic or structural order in the normal state. These characteristics make LiFeAs an attractive model system for investigating the intrinsic superconducting properties of FeSCs without the additional complexity introduced by substitutional disorder. At the same time, the absence of a nearby ordered phase makes LiFeAs a particularly stringent platform for identifying which electronic instabilities are genuinely relevant for superconductivity.

\begin{figure}[htb]
\centering
\includegraphics[width=1.0\columnwidth]{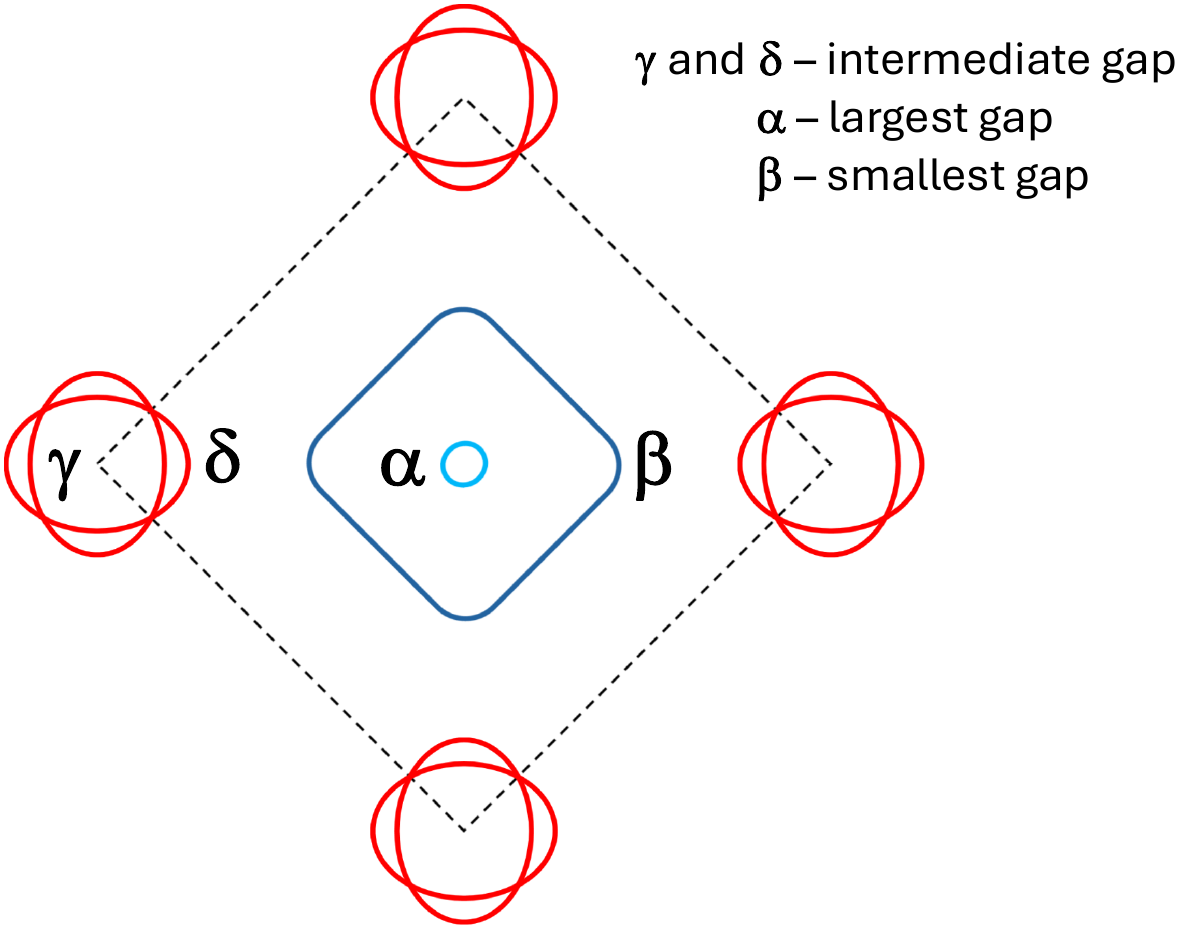}
\caption{Schematic representation of the LiFeAs Fermi surface based on ARPES results after Ref.~\cite{dai2015}. The inner hole pocket ($\alpha$) hosts the largest superconducting gap, the outer hole pocket ($\beta$) the smallest gap, while the electron pockets ($\gamma$ and $\delta$) carry intermediate gap values of nearly equal magnitude. }
\label{fig1}
\end{figure}

Despite this apparent simplicity, the superconducting state of LiFeAs remains a matter of active discussion. Angle-resolved photoemission spectroscopy (ARPES) has established a multiband Fermi surface with hole pockets around $\Gamma$ and electron pockets around $M$, together with a pronounced disparity in the superconducting gap values on the different Fermi-surface sheets \cite{borisenko2010, stockert2011, hajiri2012, dai2015, fink2019, kushnirenko2020, day2022, borisenko2016soc}. A schematic representation of the LiFeAs Fermi surface is shown in Fig.~\ref{fig1}. In the commonly used notation, the inner hole pocket ($\alpha$) possesses the largest superconducting gap, the outer hole pocket ($\beta$) the smallest one, while the two electron pockets ($\gamma$ and $\delta$) host intermediate gaps of nearly equal magnitude \cite{kushnirenko2020, umezawa2012, borisenko2012symmetry}. This band-selective gap hierarchy is important when comparing different experimental probes, because each technique weights the individual Fermi-surface sheets differently. Bulk-sensitive probes such as specific heat \cite{stockert2011, wei2010, jang2012}, lower critical field \cite{song2011, sasmal2010}, and the superfluid density study \cite{kim2011, inosov2010} are broadly consistent with nodeless multigap superconductivity, although the gap magnitudes extracted by different techniques are not identical. By contrast, tunneling \cite{chi2017, chi2012, allan2015, nag2016, sun2019, allan2012, hanaguri2012} and Andreev-reflection experiments  \cite{kuzmicheva2020, kuzmicheva2021, kuzmichev2022} often emphasize larger gap scales and in some cases suggest an effective three-gap description. Consistent with this, ARPES also supports an intrinsically three-gap superconducting structure in LiFeAs, although the corresponding gap magnitudes differ somewhat from those inferred from tunneling and Andreev-reflection experiments \cite{kushnirenko2020, umezawa2012, borisenko2012symmetry}. The resulting spread of reported energy gaps is one of the central unresolved issues for LiFeAs.

Recent spectroscopic work has also discussed the possibility of superconductivity-driven nematicity in LiFeAs \cite{sun2019, kushnirenko2020}. While such a scenario concerns rotational rather than time-reversal symmetry, it underlines the need for bulk-sensitive probes that can determine whether superconductivity is accompanied by additional broken symmetries and can isolate the intrinsic bulk condensate response.

In this context, bulk-sensitive local probes are particularly important. Zero-field (ZF) muon-spin rotation/relaxation ($\mu$SR) is among the most sensitive techniques for detecting spontaneous internal magnetic fields and therefore provides a direct test for time-reversal-symmetry breaking in the superconducting state. Transverse-field (TF) $\mu$SR, on the other hand, gives direct access to the vortex-state field distribution and thus to the magnetic penetration depth and the temperature dependence of the superfluid density. In a multiband system such as LiFeAs, this is especially valuable because $\lambda^{-2}(T)$ weights the individual bands according to their contribution to the condensate stiffness, rather than simply reflecting the largest gap scale \cite{evtushinsky2009, khasanov2009, khasanov2018, khasanov2023}. As a consequence, $\mu$SR is expected to be most sensitive to the Fermi-surface sheets that dominate the bulk superfluid response, while a weakly weighted band may remain unresolved in the measured $\rho_s(T)$. TF-$\mu$SR measurements in LiFeAs also require some care, since field-induced magnetism has been reported at elevated applied fields in samples with suppressed superconductivity \cite{pratt2009}. For this reason, the present study focuses on low applied fields in order to probe the intrinsic bulk vortex response as directly as possible.

In this work we present a combined ZF- and TF-$\mu$SR investigation of Li$_{0.95}$FeAs prepared by high-pressure self-flux growth \cite{zhigadlo2025}. The ZF measurements show no evidence for spontaneous internal fields appearing below $T_{\rm c}$ in agreement with Ref.~\onlinecite{wright2013}, thereby placing stringent constraints on time-reversal-symmetry breaking in the superconducting state. The TF data reveal a well-developed vortex response with strong pinning and a negligible nonsuperconducting contribution, demonstrating that superconductivity is a bulk property of the sample. The resulting temperature dependence of the superfluid density is well described by an effective two-gap model. By comparing the $\mu$SR results with the known multiband electronic structure of LiFeAs (see Fig.~\ref{fig1}), we show that the bulk superfluid response is dominated by the bands associated with the intermediate superconducting gaps ($\gamma$ and $\delta$) and the small gap ($\beta$), whereas the $\alpha$ Fermi-surface sheet, which hosts the largest gap, contributes only weakly to $\lambda_{ab}^{-2}$ and is therefore not resolved in the present analysis.

\section{Experimental details}

\begin{figure}[htb]
\centering
\includegraphics[width=1.0\columnwidth]{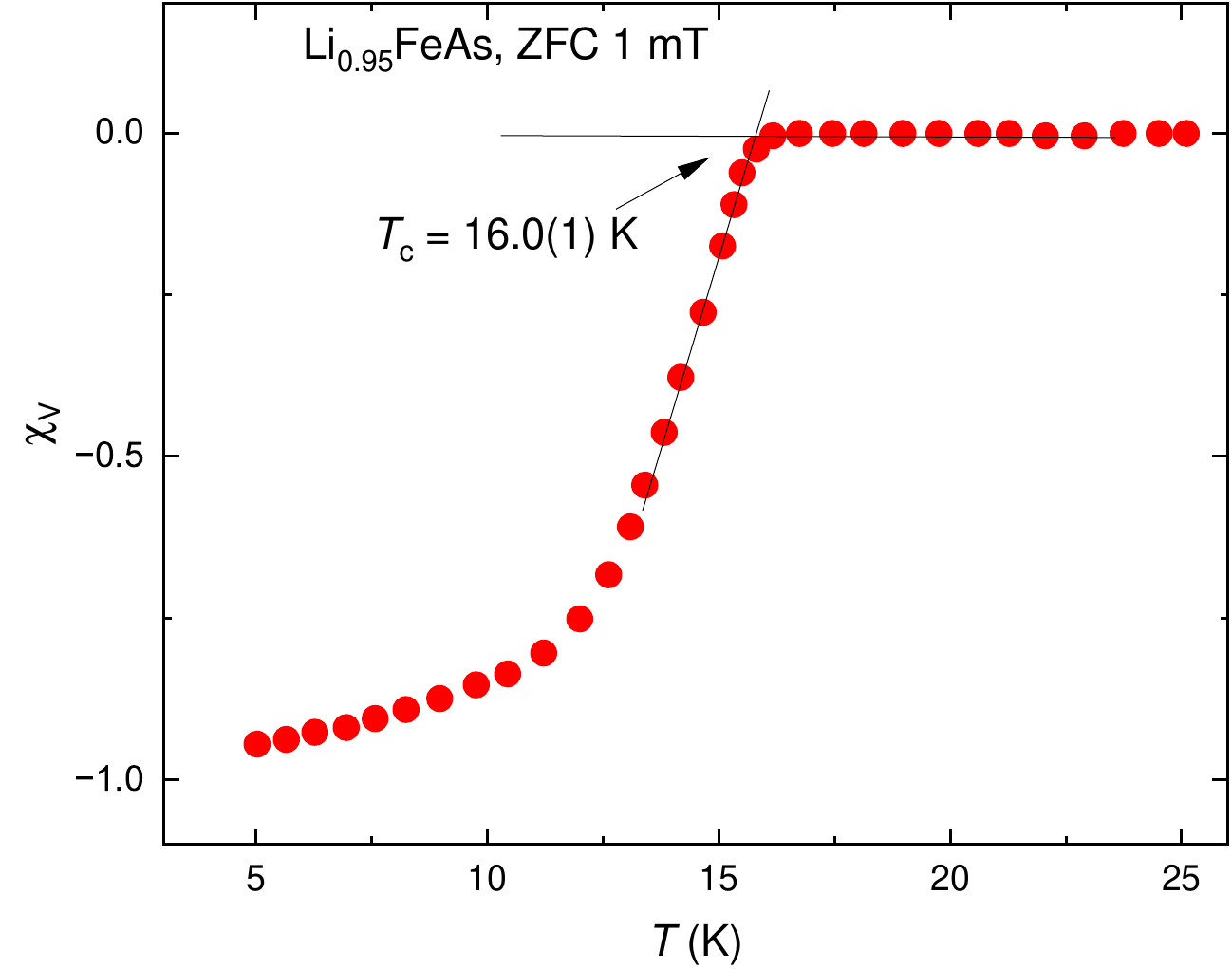}
\caption{Zero-field-cooled volume susceptibility $\chi_v(T)$ of a representative Li$_{0.95}$FeAs single crystal measured in an applied field of 1~mT. The crystal was taken from the same growth batch as that used for the $\mu$SR experiments. The straight lines indicate the procedure used to determine $T_{\rm c}=16$~K from the intersection of the linearly extrapolated transition with the $\chi_v=0$ line.}
\label{fig2}
\end{figure}

The $\mu$SR experiment was performed on a compact piece ($\sim 5$--$7$~mm in size) taken from the high-pressure-grown boule prepared as described in Ref.~\cite{zhigadlo2025}. To characterize the superconducting transition, a representative single crystal from the same boule as the $\mu$SR sample was measured in a SQUID magnetometer. The resulting zero-field-cooled volume susceptibility $\chi_v(T)$, measured in an applied field of 1~mT, is shown in Fig.~\ref{fig2}. The superconducting transition temperature, determined from the intersection of the linearly extrapolated transition with the $\chi_v=0$ line, is $T_{\rm c}\simeq16.0$~K. The low-temperature susceptibility approaches $-1$, consistent with bulk superconductivity.

The $\mu$SR measurements were carried out on the Dolly spectrometer at the $\pi$E1 beamline of the Paul Scherrer Institute (Villigen, Switzerland) using a $^4$He bath cryostat with a base temperature of about 1.5~K. Two complementary types of experiment were performed: zero-field (ZF) and transverse-field (TF) $\mu$SR. For the TF measurements, the external field was applied perpendicular to the initial muon-spin polarization, and the data were collected after field cooling in applied fields of $B_{\rm ap}=5$ and 10~mT. The $\mu$SR data were analyzed using the MUSRFIT package \cite{MUSRFIT}.
It should be noted that the sample used in the $\mu$SR experiments was a part of an as-grown boule consisting of stacked single crystals with different orientations. Therefore, as far as the $\mu$SR response is concerned, the measurements correspond effectively to those on a polycrystalline sample.

\section{Results}

\subsection{Zero-field $\mu$SR}

\begin{figure}[htb]
\centering
\includegraphics[width=0.9\columnwidth]{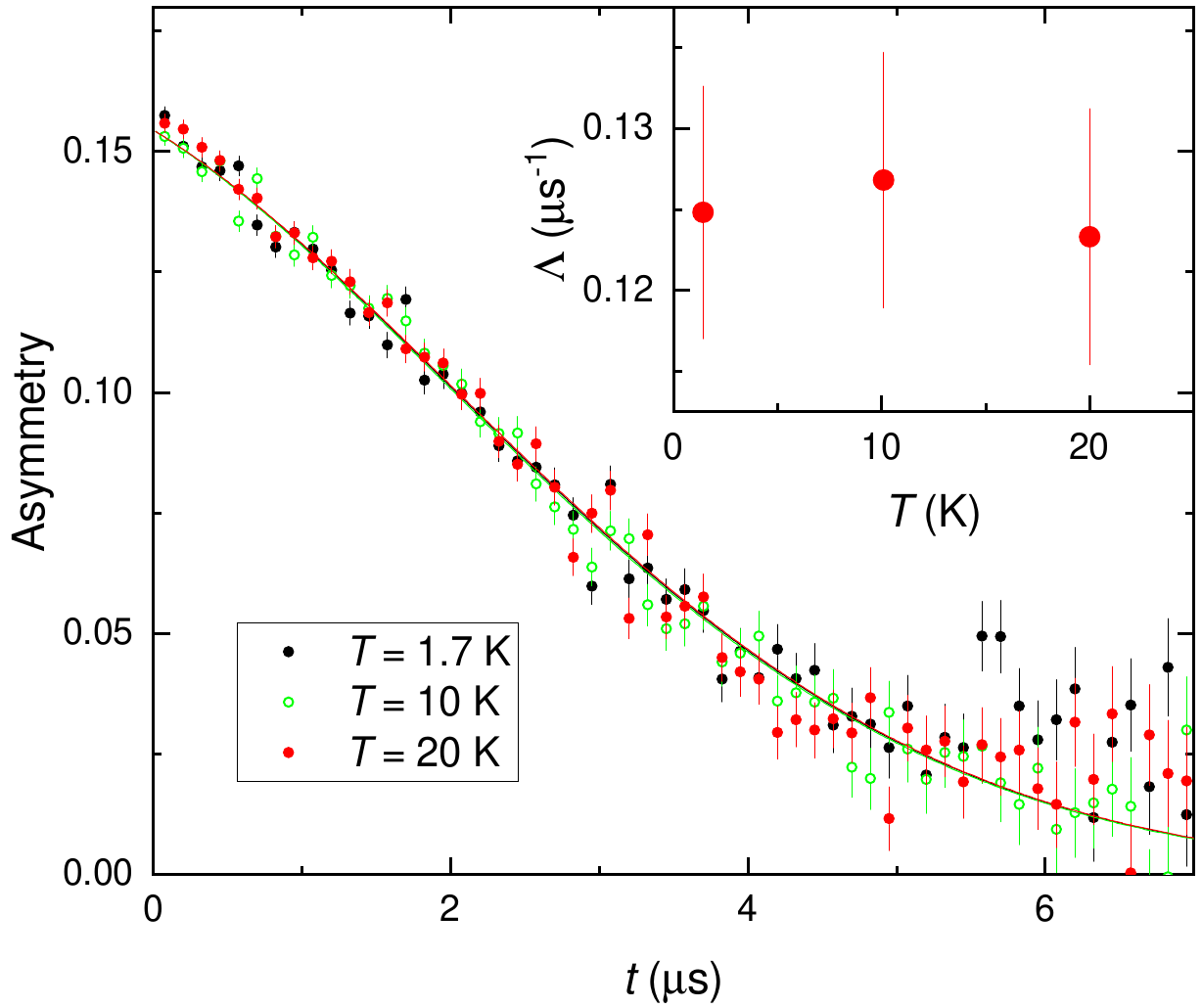}
\caption{ZF-$\mu$SR spectra of Li$_{0.95}$FeAs measured at 1.7, 10, and 20~K. The solid lines are fits to Eq.~(\ref{eq:ZF-muSR}). The inset shows the temperature dependence of the exponential relaxation rate $\Lambda$. No additional relaxation appears below $T_{\rm c}$, demonstrating the absence of a detectable spontaneous field in the superconducting state. In order to reduce correlations between the fit parameters, $\sigma_{\rm GKT}$ was assumed to be temperature independent throughout the analysis.}
\label{fig3}
\end{figure}

The ZF-$\mu$SR time spectra collected at $T=1.7$, 10, and 20~K are shown in Fig.~\ref{fig3}. The spectra exhibit only a slow relaxation and are well described by
\begin{equation}
A_{\rm ZF}(t)=A_0 \; {\rm GKT}(t) \; e^{-\Lambda t},
\label{eq:ZF-muSR}
\end{equation}
where $A_0$ is the initial asymmetry of the muon-spin ensemble, ${\rm GKT}(t)=\frac{1}{3}+\frac{2}{3}\left(1-\sigma_{\rm GKT}^2 t^2\right)e^{-\sigma_{\rm GKT}^2 t^2/2}$
is the Gaussian Kubo--Toyabe relaxation function associated with quasistatic nuclear dipolar fields ($\sigma_{\rm GKT}$ is the corresponding Gaussian relaxation rate), and $\Lambda$ is an additional exponential relaxation rate accounting for a weak electronic contribution. Note that, strictly speaking, $\Lambda$ may also contain a nuclear contribution, arising from deviations from an ideal Gaussian field distribution, for example due to the presence of isotopes with low natural abundance (such as $^{57}$Fe) that create a dilute distribution of nuclear moments. Such a combination of a static Gaussian Kubo--Toyabe term and a small exponential relaxation is commonly used in ZF-$\mu$SR when the dominant depolarization arises from nuclear moments, while residual electronic moments, dilute defects, or non-Gaussian nuclear field distributions contribute an additional Lorentzian relaxation channel \cite{Amato-Morenzoni_book_2024, Yaouanc_book_2011, Blundell_book_2022, Schenck_book_1985}.

Within the fitting uncertainty, neither the overall asymmetry nor the Lorenzian and Gaussian relaxations show any detectable anomaly on crossing $T_{\rm c}$, indicating that the local field distribution associated with the nuclear and electronic contributions remain unchanged in the superconducting state. In particular, the spectra do not exhibit any oscillatory signal or fast-relaxing component that would point to the onset of static magnetism or to the appearance of an additional magnetically distinct volume fraction below $T_{\rm c}$.

The key result is summarized in the inset of Fig.~\ref{fig3}: within the experimental uncertainty, $\Lambda$ remains unchanged above and below the superconducting transition. To reduce correlations between the fit parameters, $\sigma_{\rm GKT}$ was assumed to be temperature independent in the analysis presented in Fig.~\ref{fig3}. Thus, no additional relaxation sets in below $T_{\rm c}$, and there is no evidence for a spontaneous internal field that would be expected for a superconducting state with broken time-reversal symmetry. The ZF-$\mu$SR data therefore indicate that superconductivity in Li$_{0.95}$FeAs is time-reversal invariant within the sensitivity of the present experiment. These results are consistent with Ref.~\onlinecite{wright2013}, in which no signatures of time-reversal symmetry breaking were found in stoichiometric LiFeAs.

\subsection{Pinning and superconducting volume fraction}

\begin{figure}[htb]
\centering
\includegraphics[width=1.0\columnwidth]{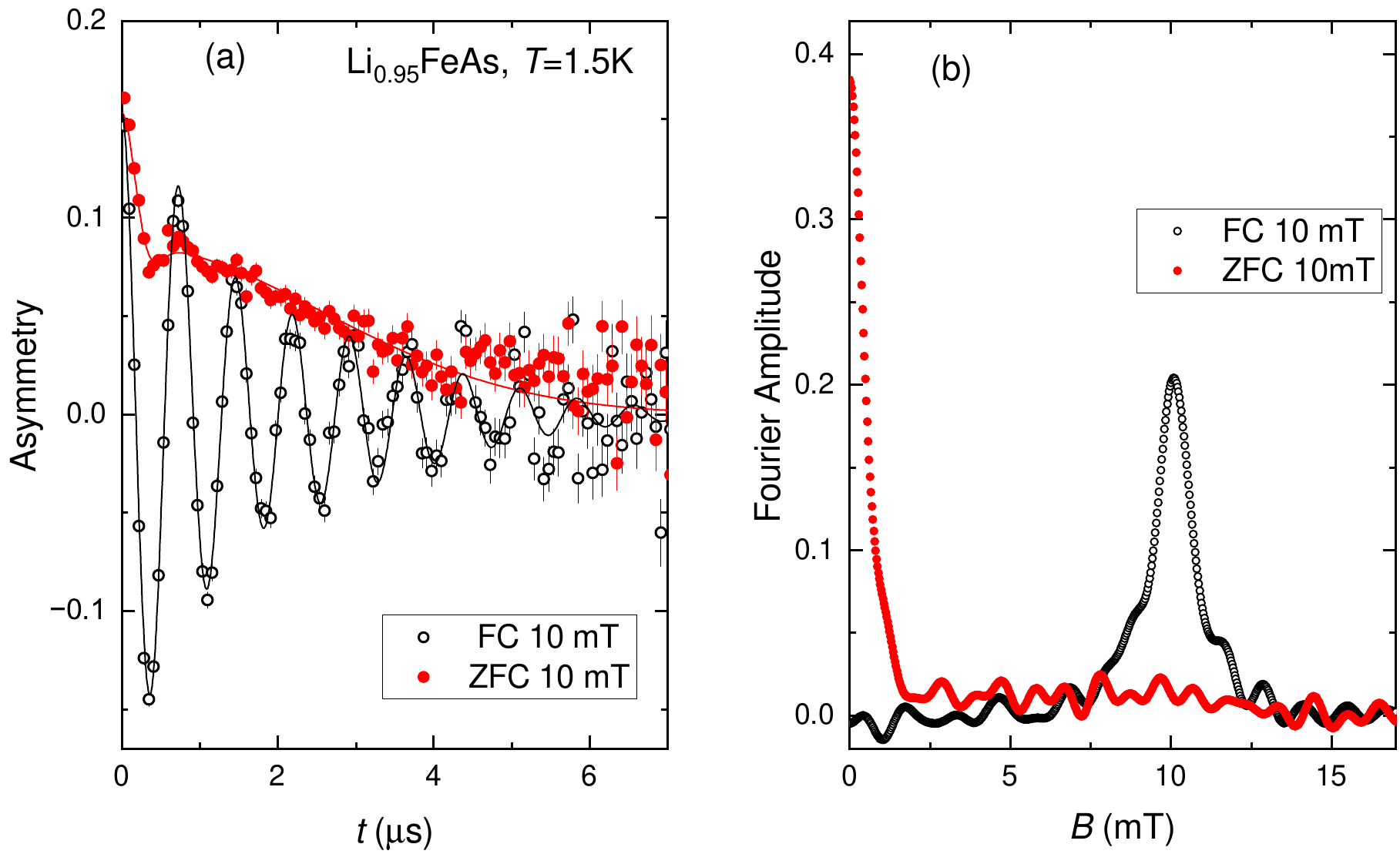}
\caption{(a) TF-$\mu$SR spectra measured at 1.5~K in an applied field of 10~mT after field cooling (open black circles) and after zero-field cooling followed by field application (red circles). The solid lines are fits to Eq.~(\ref{eq:TF}). (b) Fourier transforms of the $\mu$SR spectra shown in panel (a). The strong suppression of the 10~mT component after ZFC demonstrates pronounced flux pinning and confirms that the measured response is dominated by the superconducting phase of the sample. }
\label{fig4}
\end{figure}

Before analyzing the TF data quantitatively, it is important to assess the superconducting volume fraction and the instrumental background. For this purpose, we compared spectra recorded after zero-field cooling (ZFC) and field cooling (FC). Figure~\ref{fig4} shows the $\mu$SR time spectra [panel (a)] and the corresponding Fourier transforms [panel (b)] measured at 1.5~K in an applied field of 10~mT.

After field cooling, the field distribution is centered close to the externally applied field, as expected for a well-developed vortex-state response. By contrast, after cooling in zero field and subsequently applying the field, most of the spectral weight remains close to zero field. This behavior is characteristic of strong flux pinning: magnetic flux penetrates only a relatively thin region near the sample surface, while the main part of the sample interior remains in the Meissner state with internal field $B_{\rm int}\simeq 0$. Equally important, the ZFC field distribution does not show a pronounced component at $B \simeq B_{\rm ap}$. This demonstrates that (i) the background signal from muons missing the sample is negligible and (ii) the measured response is dominated by the superconducting phase of the sample. In other words, the TF-$\mu$SR signal is not consistent with a substantial nonsuperconducting volume fraction, confirming that superconductivity in the studied Li$_{0.95}$FeAs sample is a bulk property.

\subsection{Transverse-field $\mu$SR and magnetic penetration depth}

The temperature dependence of the superfluid density was determined from TF-$\mu$SR measurements performed after field cooling in applied fields of $B_{\rm ap}=5$ and 10~mT. These relatively low fields were chosen deliberately in order to minimize the influence of the field-induced magnetic response reported previously in LiFeAs at higher applied fields \cite{pratt2009}.

\begin{figure}[htb]
\centering
\includegraphics[width=1.0\columnwidth]{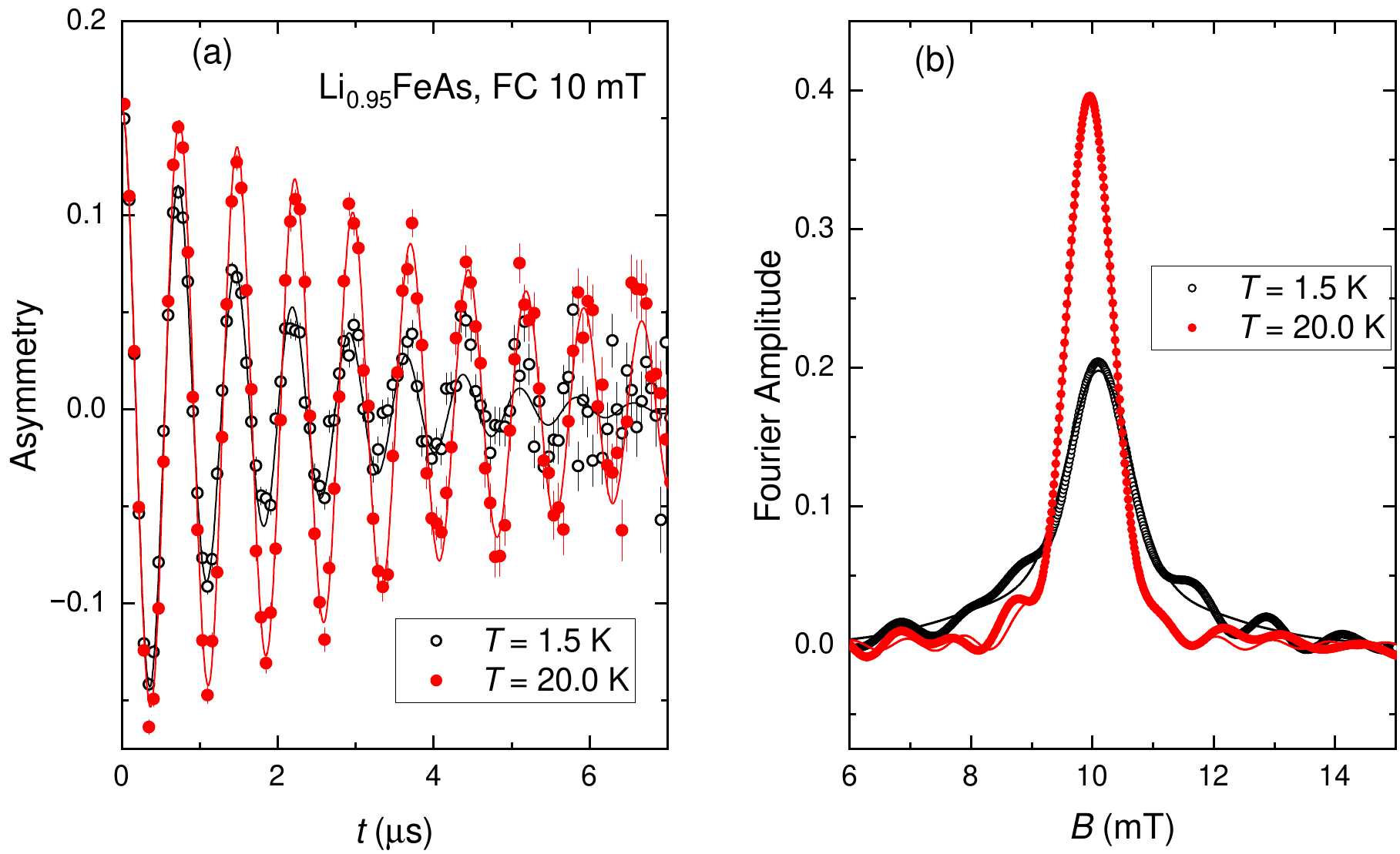}
\caption{(a) TF-$\mu$SR spectra of Li$_{0.95}$FeAs measured after field cooling in an applied field of 10~mT above ($T=20$~K, red circles) and below ($T=1.5$~K, open black circles) the superconducting transition temperature ($T_{\rm c}=16.0$~K). (b) Fourier transforms of the $\mu$SR spectra shown in panel (a). The additional broadening below $T_{\rm c}$ reflects the formation of the flux-line lattice. The solid lines are fits to Eq.~(\ref{eq:TF}).}
\label{fig5}
\end{figure}

A representative example of the TF-$\mu$SR spectra measured above ($T=20$~K) and below ($T=1.5$~K) the superconducting transition at $B_{\rm ap}=10$~mT is shown in Fig.~\ref{fig5}. On cooling below $T_{\rm c}$, the field distribution broadens owing to the formation of the vortex state. In the mixed state of an extreme type-II superconductor, this additional broadening is directly related to the inverse square of the magnetic penetration depth.

Following the standard phenomenological two-Gaussian description commonly used for moderately asymmetric vortex-state line shapes, the TF-$\mu$SR data were analyzed with \cite{maisuradze2009, khasanov2005rbos, khasanov2007}
\begin{equation}
A_{\rm TF}(t)=\sum_{i=1}^2 A_i\exp\left(-\frac{\sigma_i^2 t^2}{2}\right)\cos\left(\gamma_\mu B_i t+\phi\right),
\label{eq:TF}
\end{equation}
where $A_i$, $\sigma_i$, and $B_i$ are the asymmetry, Gaussian relaxation rate, and mean field of the $i$th component, respectively, $\gamma_\mu = 2\pi\times 135.5342$~MHz~T$^{-1}$ is the muon gyromagnetic ratio, and $\phi$ is the initial phase of the muon-spin ensemble. The corresponding field distribution is
\begin{equation}
P(B)=\sum_{i=1}^2 \frac{\gamma_\mu A_i}{\sigma_i}\exp\left[-\frac{\gamma_\mu^2(B-B_i)^2}{2\sigma_i^2}\right].\nonumber
\end{equation}
From the fitted parameters, the first and second central moments of the $P(B)$ distribution are obtained as \cite{maisuradze2009, khasanov2005rbos}
\begin{equation}
\langle B \rangle=\frac{\sum_{i=1}^2 A_i B_i}{\sum_{i=1}^2 A_i}, \nonumber
\end{equation}
and
\begin{equation}
\langle \Delta B^2 \rangle=\frac{\sum_{i=1}^2 A_i\left[\sigma_i^2/\gamma_\mu^2+(B_i-\langle B\rangle)^2\right]}{\sum_{i=1}^2 A_i}. \nonumber
\end{equation}
The fits were performed globally in the time domain using Eq.~(\ref{eq:TF}), with the asymmetries $A_1$ and $A_2$ constrained to be temperature independent, while the internal fields $B_i$ and Gaussian relaxation rates $\sigma_i$ were allowed to vary. The solid lines in Figs.~\ref{fig4} and \ref{fig5} represent the resulting fits in both the time and field domains.

\begin{figure}[htb]
\centering
\includegraphics[width=0.9\columnwidth]{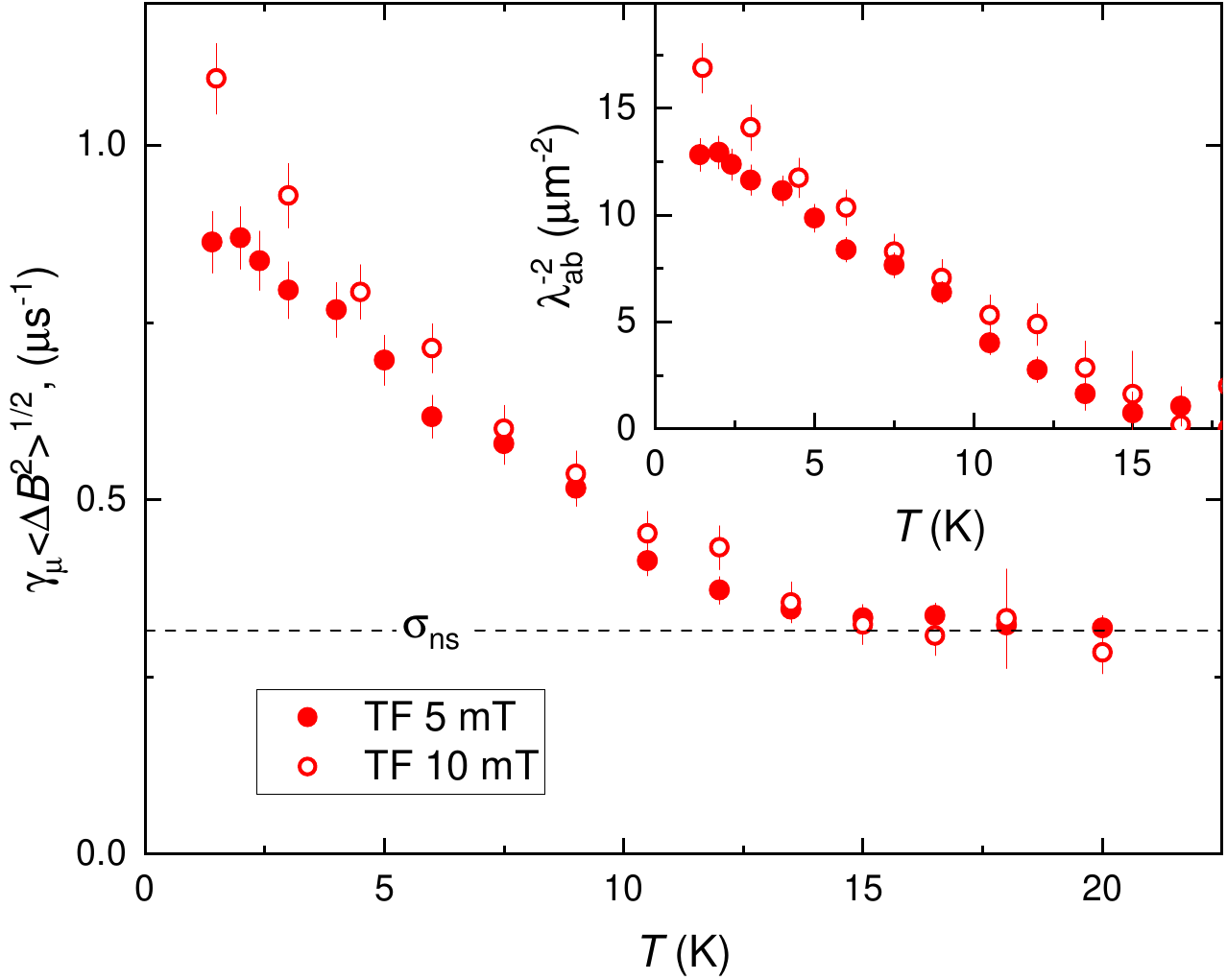}
\caption{Temperature dependence of the square root of the second central moment of the TF-$\mu$SR field distribution, $\langle \Delta B^2 \rangle^{1/2}$, for $B_{\rm ap}=5$ and 10~mT. The dashed line marks the temperature-independent normal-state contribution $\sigma_{\rm ns}$. Inset: temperature dependences of $\lambda_{ab}^{-2}(T)$ derived from Eq.~(\ref{eq:lambda_ab}).}
\label{fig6}
\end{figure}

Figure~\ref{fig6} summarizes the temperature dependence of $\langle \Delta B^2 \rangle^{1/2}$ measured at 5 and 10~mT. The normal-state contribution to the second moment, $\sigma_{\rm ns}$, was determined from the data above $T_{\rm c}$ and assumed to be temperature independent. The superconducting contribution was then obtained as \cite{Amato-Morenzoni_book_2024, Yaouanc_book_2011, Blundell_book_2022, maisuradze2009, khasanov2005rbos}
\begin{equation}
\langle \Delta B^2 \rangle_{\rm sc}=\langle \Delta B^2 \rangle-\sigma_{\rm ns}^2/\gamma_\mu^2. \nonumber
\end{equation}
For a strongly anisotropic sample with random crystallite orientation, as in the present case, the in-plane magnetic penetration depth can be estimated from \cite{brandt1988, brandt2003, sonier2000, fesenko1991}
\begin{equation}
\langle \Delta B^2 \rangle_{\rm sc}=0.00371\frac{\Phi_0^2}{\lambda_{\rm eff}^4}=0.00126\frac{\Phi_0^2}{\lambda_{ab}^4},
\label{eq:lambda_ab}
\end{equation}
where $\Phi_0 = 2.068\times10^{-15}$~Wb is the magnetic flux quantum and the effective penetration depth $\lambda_{\rm eff}=1.31\lambda_{ab}$ accounts for the anisotropic powder average.

Using Eq.~(\ref{eq:lambda_ab}), we obtain $\lambda_{ab}(1.5~\mathrm{K}) = 275(15)$~nm for 5~mT and $245(15)$~nm for 10~mT. The slightly larger value found at 5~mT is not considered intrinsic; rather, it most likely reflects the fact that this field lies closer to the first critical field $H_{c1}$, where the second moment of the vortex-state field distribution is reduced \cite{pratt2009, brandt2003}. The temperature dependence of $\lambda_{ab}^{-2}$ is presented in the inset of Fig.~\ref{fig6}.

\section{Discussion}

\subsection{Constraints on time-reversal-symmetry-breaking states}

An important outcome of the present ZF-$\mu$SR measurements is that no additional relaxation appears below $T_{\rm c}$, implying that the superconducting transition in Li$_{0.95}$FeAs is not accompanied by detectable spontaneous internal magnetic fields. This places strong constraints on superconducting states with an intrinsically complex order parameter, i.e. states that break time-reversal symmetry and would be expected to generate spontaneous local magnetic fields in the bulk. In particular, this result disfavors scenarios involving mixed complex order parameters of the $s+is$ or $s+id$ type, which have been discussed theoretically for LiFeAs \cite{ahn2014}. At the same time, the absence of a ZF-$\mu$SR anomaly does not by itself distinguish between time-reversal-invariant real $s$-wave states such as $s_{++}$ and $s_{\pm}$.

This distinction is important in the context of the ongoing discussion of the superconducting order parameter in LiFeAs. On the one hand, theoretical analysis of the multiband pairing problem has shown that several different real $s$-wave configurations are possible in LiFeAs, and that under appropriate conditions the system may also enter a mixed state with broken time-reversal symmetry at lower temperature \cite{ahn2014}. On the other hand, phase-sensitive Bogoliubov quasiparticle-interference experiments have been interpreted in favor of a sign-changing $s_{\pm}$ gap structure, with opposite signs of the superconducting order parameter on the hole and electron bands \cite{chi2014,altenfeld2018}. The present ZF-$\mu$SR data therefore do not select uniquely between different real $s$-wave states, but they do provide an important bulk constraint by ruling out, within the sensitivity of the experiment, a superconducting state accompanied by spontaneous magnetic fields.

It is also useful to view the present result in the context of earlier proposals of triplet pairing in LiFeAs. An earlier ZF-$\mu$SR study of stoichiometric LiFeAs likewise found no evidence for time-reversal-symmetry breaking and concluded that the data do not support triplet-pairing scenarios requiring spontaneous internal fields \cite{wright2013}. In particular, spin-triplet $p$-wave superconductivity driven by nearly ferromagnetic fluctuations has been proposed for LiFeAs \cite{brydon2011}, while a chiral triplet $p_x+ip_y$ state was discussed for slight deviations from stoichiometry \cite{aperis2013}. Although a null ZF-$\mu$SR result does not exclude every possible triplet state, it strongly disfavors chiral or otherwise time-reversal symmetry breaking triplet states that would be expected to generate spontaneous magnetic fields. In this respect, the present $\mu$SR data are consistent with neutron-scattering results indicating predominantly antiferromagnetic, rather than ferromagnetic, spin fluctuations in LiFeAs \cite{taylor2011}.

\subsection{Bulk weighting of the multigap superfluid response}

Information on the superconducting gap structure can be obtained from the temperature evolution of the superfluid density. In the present case, the latter is directly accessible through the measured in-plane magnetic penetration depth, since $\rho_s(T)\propto \lambda_{ab}^{-2}(T)$. By normalizing the experimental $\lambda_{ab}^{-2}(T)$ data to their zero-temperature value $\lambda_{ab}^{-2}(0)$, one obtains the normalized superfluid density $\rho_s(T)=\lambda_{ab}^{-2}(T)/\lambda_{ab}^{-2}(0)$, which is particularly suitable for analyzing the superconducting gap structure.

The normalized superfluid density derived from the TF-$\mu$SR data is shown in Fig.~\ref{fig1}. Its temperature dependence is well described by an effective two-gap model \cite{khasanov2007, khasanov2009, niedermayer2002},
\begin{equation}
\rho_s(T)=\omega\,\rho_{s,1}(T,\Delta_1)+(1-\omega)\,\rho_{s,2}(T,\Delta_2),
\label{eq:rho_s}
\end{equation}
with the individual components given by \cite{khasanov2005rbos, tinkham1996, carrington2003}
\begin{equation}
\rho_{s,i}(T)=1+2\int_{\Delta_i(T)}^{\infty}\left(\frac{\partial f}{\partial E}\right)\frac{E\,dE}{\sqrt{E^2-\Delta_i(T)^2}}, \nonumber
\label{eq:rho_single}
\end{equation}
where $f=[1+\exp(E/k_{\rm B}T)]^{-1}$ is the Fermi function and the temperature dependence of the gap is approximated by \cite{carrington2003, khasanov2020beau}
\begin{equation}
\Delta_i(T)=\Delta_i \tanh\left\{1.82\left[1.018\left(\frac{T_{\rm c}}{T}-1\right)\right]^{0.51}\right\}. \nonumber
\label{eq:DeltaT}
\end{equation}
The fit yields $\Delta_1=2.0(2)$~meV, $\Delta_2=0.7(2)$~meV, and $\omega=0.61(2)$.

These values are close to those inferred from bulk-sensitive probes such as specific heat and lower-critical-field measurements \cite{stockert2011,wei2010, jang2012, song2011, sasmal2010}. They differ more substantially from some ARPES \cite{kushnirenko2020, umezawa2012, borisenko2012symmetry}, tunneling  \cite{chi2017, chi2012, allan2015, nag2016, sun2019, allan2012, hanaguri2012}, and Andreev-reflection results \cite{kuzmicheva2020, kuzmicheva2021, kuzmichev2022} that emphasize larger gap scales and, in several cases, resolve three distinct gap values. This difference is not unexpected in a multiband material such as LiFeAs. The superconducting condensate is distributed over several Fermi-surface sheets, and each experimental technique probes a different weighted average over momentum space.

\begin{figure}[htb]
\centering
\includegraphics[width=0.9\columnwidth]{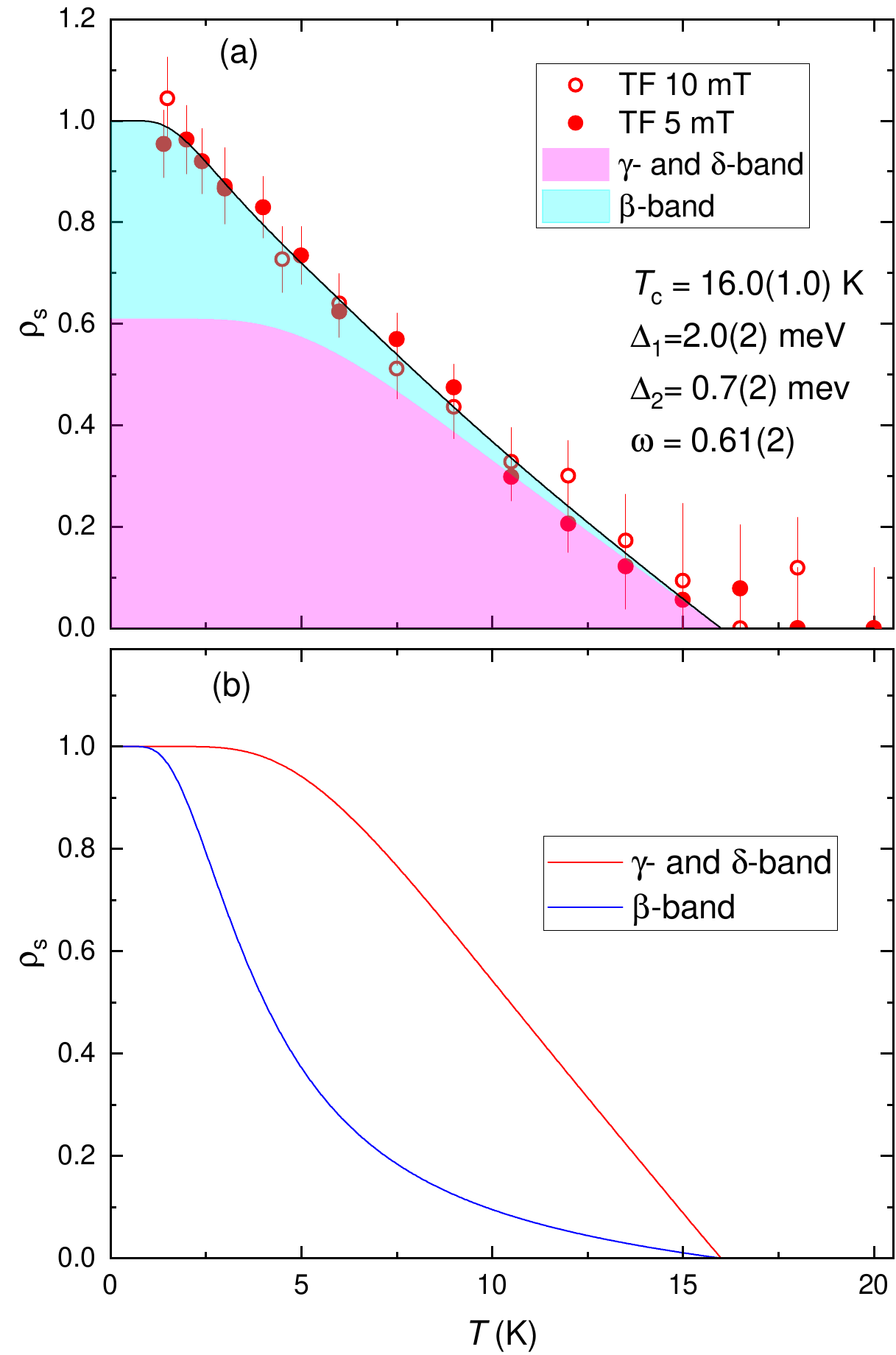}
\caption{(a) Temperature dependence of the normalized superfluid density $\rho_s(T)=\lambda_{ab}^{-2}(T)/\lambda_{ab}^{-2}(0)$ for Li$_{0.95}$FeAs. The solid line is a fit to the effective two-gap model of Eq.~(\ref{eq:rho_s}). The colored areas illustrate the decomposition into two effective superfluid-density components. The closed and open symbols correspond to the results of the 5~mT and 10~mT TF-$\mu$SR measurements, respectively. (b) Partial superfluid-density contributions obtained from the fit.}
\label{fig7}
\end{figure}

A useful way to rationalize this point is to recall that the zero-temperature contribution of the $i$th band to $\lambda_{ab}^{-2}(0)$ is proportional to the integral of the Fermi velocity around the corresponding Fermi-surface sheet \cite{evtushinsky2009, khasanov2009, chandrasekhar1993}:
\begin{equation}
I_i \propto \oint_{i{\rm th\ band}} v_{{\rm F},i}(\mathbf{k})\,dk.
\label{eq:Lambda0}
\end{equation}
Using representative ARPES parameters for LiFeAs \cite{kushnirenko2020,borisenko2012symmetry}, one obtains the approximate band weights listed in Tab.~\ref{tab1}. The important point is that the inner hole pocket ($\alpha$), on which ARPES finds the largest superconducting gap, contributes only a very small fraction to the total superfluid density, as shown in the last column of Tab.~\ref{tab1}. By contrast, the $\beta$, $\gamma$, and $\delta$ sheets dominate the bulk response. Thus, a bulk probe such as $\mu$SR naturally gives greater weight to the intermediate and smaller gap scales, whereas surface-sensitive spectroscopies can more readily resolve the large gap on the weakly weighted $\alpha$ pocket.

\begin{table}[t]
\caption{Representative parameters extracted from ARPES data for LiFeAs, including the mean Fermi velocity $\langle v_{\rm F}\rangle_i$, the average Fermi-surface diameter $\langle d_{\rm F}\rangle_i$, and the relative contribution $I_i/\sum I_i$ of the $i$th band to the superfluid density. The values of $\langle v_{\rm F}\rangle_i$ and $\langle d_{\rm F}\rangle_i$ are representative of ARPES-based estimates compiled from the literature \cite{borisenko2010, borisenko2012symmetry}. The values of $I_i/\sum I_i$  are derived by Eq.~(\ref{eq:Lambda0}). }
\label{tab1}
\centering
\begin{tabular}{l|ccc}
     &  &        &\\
Band\; & $\langle v_{\rm F}\rangle_i$ & $\langle d_{\rm F}\rangle_i$ & $I_i/\sum I_i$ \\
     &   (eV nm)                    &   (nm$^{-1}$)                &\\
\hline
$\alpha$ & $\sim 0.01$  & $\sim 1.0$  & 0.03 \\
$\beta$  & 0.022  & 4.0  & 0.30 \\
$\gamma$ & 0.035  & 3.0  & 0.31 \\
$\delta$ & 0.035  & 2.6  & 0.36 \\
\end{tabular}
\end{table}

In this sense, the present two-gap fit should be regarded as an effective description of the bulk superfluid density rather than as evidence that only two bands are superconducting. Bands with similar gap magnitudes can be combined into an effective contribution, while the weakly weighted large-gap contribution has only a minor influence on the overall temperature dependence of $\lambda_{ab}^{-2}(T)$. This interpretation explains why the present $\mu$SR analysis remains fully compatible with the multiband electronic structure established by ARPES.

A further important point is that the relative band weights inferred from the $\mu$SR analysis are in reasonably good agreement with the ARPES-based estimate. Experimentally, the effective two-gap fit yields contributions of about 0.61 and 0.39 for the larger- and smaller-gap components, respectively (see Fig.~\ref{fig1} and the discussion above). By contrast, the band-structure estimate based on ARPES gives a total weight of about 0.7 for the $\gamma$ and $\delta$ sheets, on which the intermediate superconducting gap opens, and about 0.3 for the $\beta$ sheet carrying the smallest gap (see Tab.~\ref{tab1} and Fig.~\ref{fig1}). Given that these values are obtained by two very different methods, the level of agreement is remarkably good.
This consistency strongly supports the conclusion that the $\mu$SR response is dominated by the $\gamma$, $\delta$, and $\beta$ sheets, whereas the $\alpha$ sheet, despite hosting the largest superconducting gap, contributes too weakly to the total superfluid density to be resolved in the present analysis.

\section{Conclusions}

In summary, we have carried out a combined ZF- and TF-$\mu$SR study of Li$_{0.95}$FeAs grown by a high-pressure self-flux method. Complementary magnetization measurements confirm bulk superconductivity with $T_{\rm c}\simeq 16$~K. The ZF-$\mu$SR data collected in the temperature range 1.7--20~K show no detectable enhancement of the electronic relaxation rate below $T_{\rm c}$, providing no evidence for spontaneous internal fields and thereby placing strong constraints on time-reversal-symmetry breaking in the superconducting state.

The TF-$\mu$SR measurements performed in applied fields of 5 and 10~mT reveal a well-developed vortex response with strong flux pinning and only a negligible nonsuperconducting contribution, confirming that the measured signal is dominated by the bulk superconducting state. From the temperature-dependent second moment of the internal field distribution we obtain $\lambda_{ab}(1.5~\mathrm{K}) = 245(15)$~nm. The temperature dependence of the normalized superfluid density is well described by an effective two-gap model with $\Delta_1 = 2.0(2)$~meV and $\Delta_2 = 0.7(2)$~meV.

These results establish Li$_{0.95}$FeAs as a bulk multigap superconductor without detectable time-reversal-symmetry breaking. Comparison with the ARPES-based multiband electronic structure further shows that the superfluid density measured by $\mu$SR is essentially insensitive to the $\alpha$ Fermi-surface sheet, on which the largest superconducting gap opens, because this sheet contributes only weakly to the total superfluid density, at the level of $\sim 3$\%. Instead, the $\mu$SR response is dominated by the sheets associated with the intermediate superconducting gaps ($\gamma$ and $\delta$) and the small gap ($\beta$). In particular, the ARPES-based band-structure estimate yields relative contributions of about 0.7 and 0.3 for the intermediate- and small-gap sheets, respectively, in good agreement with the corresponding weights obtained from the $\mu$SR analysis, 0.61 and 0.39. In this way, the present work demonstrates that $\mu$SR provides direct access to the intrinsic bulk superfluid response of Li$_{0.95}$FeAs and helps reconcile the apparent spread of superconducting gap values reported by different experimental probes.

%\ack
%The experiments were performed at muon instrument Dolly (S$\mu$S, PSI, Villigen).

\end{document}